\pdfoutput=1

\documentclass[conference,compsoc]{IEEEtran}
\ifCLASSOPTIONcompsoc
  \usepackage[nocompress]{cite}
\else
  \usepackage{cite}
\fi
\ifCLASSINFOpdf
\else
\fi

\usepackage{stfloats}
\fnbelowfloat
\usepackage{amsmath,amsfonts}
\usepackage{multirow}
\usepackage{makecell}
\usepackage{algorithmic}
\usepackage{algorithm}
\usepackage{graphicx}
\usepackage{textcomp}
\usepackage{xcolor}
\usepackage{url}
\usepackage{amsmath}
\usepackage{array}
\usepackage{booktabs}
\usepackage{fullpage}
\usepackage{graphicx}
\usepackage{threeparttable}
\usepackage{wasysym}
\usepackage{pgfplots}
\pgfplotsset{compat=newest}
\usepackage{subcaption}

\newtheorem{definition}{Definition}

\begin{document}
%
\title{DataSeal: Ensuring the Verifiability of Private Computation \\ on Encrypted Data}

\author{
\IEEEauthorblockN{Muhammad Husni Santriaji\IEEEauthorrefmark{2}\IEEEauthorrefmark{3}, Jiaqi Xue\IEEEauthorrefmark{3}, Qian Lou\IEEEauthorrefmark{3}\IEEEauthorrefmark{1}, Yan Solihin\IEEEauthorrefmark{3}\IEEEauthorrefmark{1}}
\IEEEauthorblockA{
\IEEEauthorrefmark{2}Universitas Gadjah Mada, muhammad.husnisantriaji@ugm.ac.id \\
\IEEEauthorrefmark{3}University of Central Florida, \{jiaqi.xue, qian.lou, yan.solihin\}@ucf.edu
}
}

\maketitle

\def\thefootnote{$*$}\footnotetext{Much of the research was conducted when Santriaji was a postdoctoral researcher in the ARPERS group at UCF. Lou and Solihin are corresponding authors: \{qian.lou, yan.solihin\}@ucf.edu}
\begin{abstract}
Fully Homomorphic Encryption (FHE) allows computations to be performed directly on encrypted data without needing to decrypt it first. This "encryption-in-use" feature is crucial for securely outsourcing computations in privacy-sensitive areas such as healthcare and finance. Nevertheless, in the context of FHE-based cloud computing, clients often worry about the integrity and accuracy of the outcomes. This concern arises from the potential for a malicious server or server-side vulnerabilities that could result in tampering with the data, computations, and results.  Ensuring integrity and verifiability with low overhead remains an open problem, as prior attempts have not yet achieved this goal. To tackle this challenge and ensure the verification of FHE's private computations on encrypted data, we introduce DataSeal, which combines the low overhead of the algorithm-based fault tolerance (ABFT) technique with the confidentiality of FHE, offering high efficiency and verification capability.  Through thorough testing in diverse contexts,  we demonstrate that DataSeal achieves much lower overheads for providing computation verifiability for FHE than other techniques that include MAC, ZKP, and TEE. DataSeal's space and computation overheads decrease to nearly negligible as the problem size increases.  

\end{abstract}


\IEEEpeerreviewmaketitle

\section{Introduction}

\textit{Fully Homomorphic Encryption (FHE)}~\cite{gentry2009fully, brakerski2014leveled, halevi2019BFVimproved, kim2022approximate} is a cryptographic privacy-preserving computation technique that enables a unique computation model where mutually-distrusting parties can collaborate. For instance, computation providers, such as Machine Learning as a Service (MLaaS) model operators or financial analysis model proprietors, aim to offer services to clients without revealing their proprietary models. Simultaneously, clients seek to supply data for computations while keeping the contents confidential. 
FHE facilitates this cooperative model by allowing the server to process the client's encrypted data directly without first decrypting it. This ensures the confidentiality of both the client's data and the output. Therefore, FHE acts as a dual-purpose safeguard, securing sensitive client data as well as proprietary models, making it a crucial tool for enabling privacy-protected computations in cloud environments~\cite{AmazonScience, IBMSecurity, GoogleCloud, Google, Microsoft, reagen2021cheetah}.

FHE has seen a surge of interest in the computing industry in recent years. FHE has been deployed in learning-enabled systems to build privacy-preserving machine learning inferences, deployed on real-world products including Duality Techniques~\cite{DualityTech}, Cape Privacy~\cite{capeprivacy}, Zama~\cite{Zama}, and Inpher~\cite{Inpher}. Some of FHE's weaknesses are being addressed. For example, its high execution time has improved substantially and over the past two years, the latency for deep neural network inference on ResNet-20 has decreased from 88,320 seconds to just 15.9 seconds~\cite{fan2023tensorfhe,samardzic2022craterlake, kim2022ark}, a phenomenal progress which will likely continue.

While FHE ensures data privacy, it ensures neither {\em computation integrity} (the property that code and data are not tampered with as they are used) nor {\em verifiability} (the property that the correct computation was performed). Verifiability includes computation integrity but adds the requirement that the code executed implements the correct computation. It is critical for a client to be able to verify the correctness of the calculation performed by the server. An adversary may compromise service to modify computation or user data,  leading to wrong output. Alternatively, an adversary may compromise the system (OS or hypervisor) or even indirectly modify data in memory without code compromise, e.g., via a RowHammer attack. This means that the client may unknowingly receive the wrong computation results. 

Recognizing this limitation, a cryptographic approach for integrity-checking protocols has been developed to address these integrity challenges and complement FHE. These include homomorphic message authentication codes (MAC, as discussed in \cite{chatel2022verifiable}) and zero-knowledge proofs (ZKP, as outlined in \cite{viand2023verifiable}). While these methods assure the integrity of computations on encrypted data, they do not ensure verifiability independently and introduce significant computational overhead due to their complexity regarding computational requirements and data volume. This leads to substantial processing delays and increased resource demands, posing a challenge for practical implementation ~\cite{bhadauria2020ligero++, bunz2018bulletproofs, gennaro2010non, goldwasser2015delegating, parno2016pinocchio, weng2021wolverine, brakerski2011fully, bois2021flexible, fiore2014efficiently, fiore2020boosting, ganesh2021rinocchio}.

The computation providers could also run their models in Trusted Execution Environments (TEEs). However, relying on TEEs is problematic for several reasons. First, TEEs do not inherently ensure verifiability, as the client cannot automatically verify if the correct computation was performed by the TEE~\footnote{An attestation can solve this but it requires the client to trust the compute platform correctness, the attester, and the correct thing is being attested.}  Second, TEE support is not universal; platforms such as NPUs and many GPUs do not support TEEs. Finally, some TEEs may not support full integrity protection even when TEE is supported. For example, AMD SEV~\cite{AMD_SEV} and Intel Scalable SGX~\cite{Intel_SGX} only protect the integrity of one enclave against other enclaves but not against direct memory tampering incurred by, say, RowHammer attacks. Only Intel SGX supports full integrity protection via an integrity tree~\cite{intel2021sgx}. 

Algorithm-Based Fault Tolerance (ABFT) is a method developed to verify computation results by embedding redundancy within data structures. Primarily used to detect and correct errors in high-performance computing, ABFT employs checksum techniques in matrix operations by adding extra rows or columns \cite{abft,abft1,abft2,ABFT-1984-checksum}. These checksums adapt during computations to identify discrepancies from faults. While effective for hardware failures or unintentional errors, ABFT is not designed to ensure data confidentiality or protect against intelligent adversaries that may manipulate data and checksum simultaneously to evade detection. 

In this paper, we propose \textit{DataSeal}, an approach that enables the \textit{verifiability} of FHE computations through the strategic integration of ABFT with low overheads. The core innovation of DataSeal lies in incorporating ABFT into the FHE framework by embedding secrets and redundancy within the original data. 
Similar to ABFT, the client can, at low cost, verify that the correct computation was performed by analyzing the redundancy in the resulting output produced by the cloud. Unlike ABFT, the redundant data is calculated using client-specific secret key, hence not having the key, the attacker cannot produce a fake checksum that evades verification. This synergy allows for the execution of FHE computations while allowing the verification of their computation results. The redundant data is also encrypted with FHE, ensuring confidentiality.  

To summarize, our contributions are:
\begin{enumerate}
\item A new approach (referred to as {\em DataSeal}), where we adapted ABFT for integration into FHE to enable computation verification while fortifying its security with a client-private key.  
\item The design of DataSeal for various matrix computations with low space overheads (i.e., two additional rows are added regardless of the matrix size), and scalable computation overheads (i.e. diminishing verification overheads as the matrix size increases). 
\item We demonstrate that DataSeal is secure against targeted data modifications. 
\end{enumerate}

We conduct performance evaluations for individual matrix operations (addition, multiplication, and convolution) and end-to-end DNN inference (including activation) to demonstrate the efficacy and practicality of our approach. The evaluation looks into the performance overheads on the server side. The evaluation of DataSeal shows its substantially lower overheads than other methods such as MAC, ZKP, and TEE, sometimes by one order of magnitude or higher. Furthermore, we found that DataSeal's verification overheads decrease relative to the computation as the problem size increases. In contrast, other schemes, such as FHE with MAC, FHE with ZKP, and FHE with TEE, their overheads tend to increase with the problem size. Thus, as the problem size increases, the performance gap between DataSeal and other schemes increases. For large enough matrix sizes, DataSeal overheads become nearly negligible.   This comparative analysis establishes DataSeal's unique attractiveness.


\section{Threat Model}
We examine a scenario where a computation \( f(x): X \rightarrow Y \) is outsourced between a client \( \mathcal{C} \), who possesses the input \( x \), and a server \( \mathcal{S} \). The ownership of the function \( f(x): X \rightarrow Y \) varies: it may reside with the client \( \mathcal{C} \), in line with client-focused frameworks like Infrastructure as a Service (IaaS) \cite{serrano2015infrastructure} or Platform as a Service (PaaS) \cite{keller2010platform}, or it could belong to the server \( \mathcal{S} \), as seen in services like Machine Learning as a Service (MLaaS) \cite{ribeiro2015mlaas}, where the server is responsible for providing the model data (e.g., the weights in a Deep Neural Network). In our model, we assume that if part of the data \( X \) belongs to the server \( \mathcal{S} \), it is known to both parties. However, data originating from the client \( \mathcal{C} \) must be kept confidential. The primary computation we focus on is matrix operations like matrix multiplication and convolution. We remark that extending DataSeal to other computations would require new ABFT techniques.

The threat model we advance considers the possibility of the server \( \mathcal{S} \) being malevolent or compromised, possessing the ability to alter ciphertexts. This extends beyond the conventional semi-honest model, which perceives \( \mathcal{S} \) as simply honest but curious (HBC) \cite{le2023privacy,dong2022fusion, badolato2022privacy}. While the HBC approach offers some promise, it is vulnerable to the threat of malign servers. Our model is designed to address proactive adversarial actions, including data manipulation and unauthorized data access. In this context, the adversarial server's objectives are twofold: (i) Violating data privacy by accessing and potentially understanding the client's data $x$; (ii) Compromising computation integrity by modifying the data in a way that changes its meaning or accuracy, all while avoiding detection. 

To ensure the security of DataSeal under this threat model, we define a formal security game that captures the server's ability to forge computation results. DataSeal's security is based on the unforgeability of the encoded matrices and checksums transmitted to the server. The adversary’s goal is to produce a forged result that passes client-side verification without access to the secret key or random scalar. As detailed in Appendix A, the probability that an adversary can successfully generate a valid forged result is shown to be negligible under standard cryptographic assumptions. Please refer to Appendix A for a detailed description of the security guarantees provided by DataSeal.


\section{Background and Related Works}
\subsection{Background}

\noindent\textbf{Algorithm-Based Fault Tolerance (ABFT).}
ABFT \cite{wang1994algorithm,chen2013online, zhao2020ft} is a method employed to add verifiability of computation on a server by embedding redundancy into data, typically for matrix-based operations. After computation, the resulting redundant result enables the client to verify if the correct result was produced in computation. Due to this property, ABFT enables a system-agnostic error correction directly within the computational algorithms, allowing fault tolerance improvement without relying on external system redundancies or hardware-based solutions. The primary advantage of ABFT is its flexibility. Its flexibility comes from it being hardware-agnostic allowing it to be used in any hardware platform, and from the overheads incurred only when used in computation. 

\begin{figure}[ht]
    \centering
\includegraphics[width=0.9\linewidth]{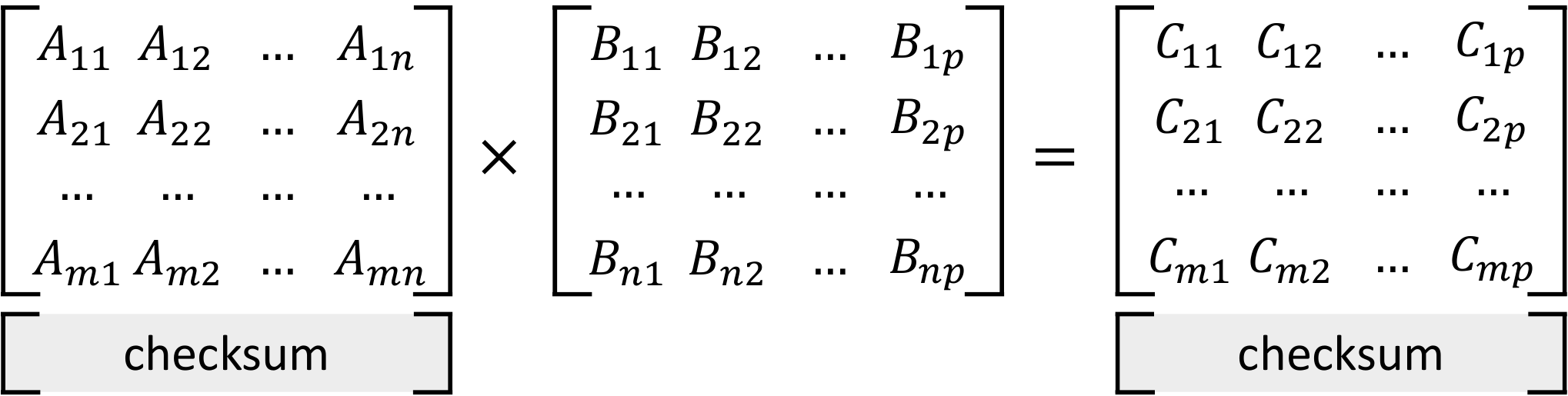}
    \caption{ABFT in Matrix Multiplication.}
    \label{fig:mat_checksum}
\end{figure}

Figure~\ref{fig:mat_checksum} illustrates the application of ABFT in matrix multiplication. It displays two matrices: matrix \( A \) with an additional checksum row at the bottom and matrix \( B \) on the left. When matrix \( A \) is multiplied by matrix \( A \), the output is matrix \( C \), which also includes a checksum row at the bottom. This checksum row in matrix \( C \) is key to verifying the multiplication process's integrity and correctness. The ABFT approach ensures that any computational errors or discrepancies can be detected through the checksums, thereby providing a reliable method for maintaining the accuracy and integrity of computations in scenarios prone to errors or faults. The inclusion of checksums in matrix operations, as depicted, demonstrates an effective use of redundancy within data structures to enhance computational reliability and security.


\begin{figure}[ht]
    \centering
    \includegraphics[width=0.7\linewidth]{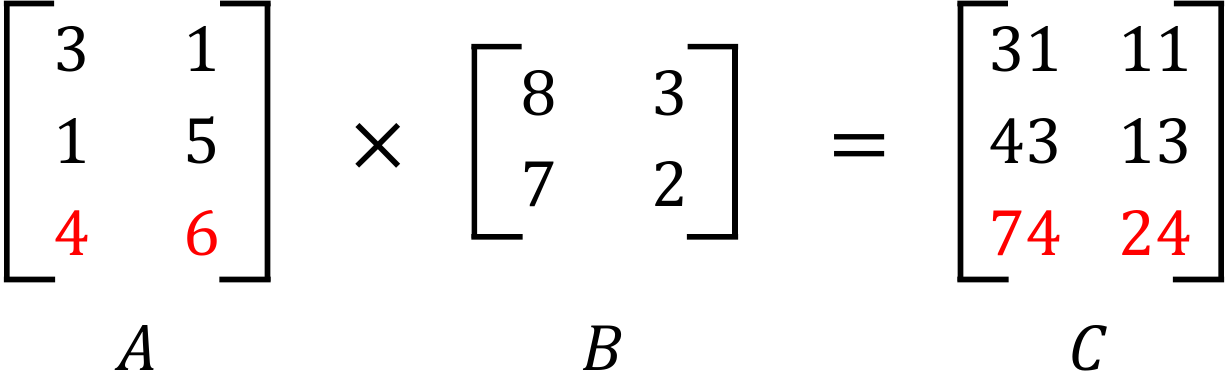}
    \caption{An example of matrix multiplication with ABFT checksums. Elements in red are redundant data added for verification.}
    \label{fig:example_ABFT}
\end{figure}


A simplified example is illustrated in Figure~\ref{fig:example_ABFT}. Matrix \(A\) is multiplied by matrix \(B\) to produce matrix \(C\), with both matrices initially \(2\times 2\). An additional row in matrix \(A\) (in red) acts as a checksum, calculated as the sum of elements in each column (e.g., \(4 = 3+1\)). After multiplication, matrix \(C\) also contains an additional row. In an error-free execution, each redundant element in \(C\) is a checksum of its column (e.g., \(74 = 31+43\)). Any error affecting memory or logic will cause a mismatch in these checksums, thus illustrating ABFT's capability to maintain data integrity and verify matrix multiplication correctness.

Note, however, while ABFT can verify computation when dealing with a random fault, its security is not strong. In the example above, if the attacker changes both values of an element and its checksum, the attacker can evade the verification. For example, the attacker may add the same constant to an element of matrix \(A\) and to the checksum, e.g. change $A_{00}$ to 4 and $A_{02}$ to 5. In the resulting matrix \(C\), $C_{00} = 39$ and $C_{02} = 82$, which passes checksum verification and the attack goes undetected. Finally, while ABFT can verify a simple matrix operation (i.e. multiplication), it has not been demonstrated that it can deal with end-to-end DNN inference.

\noindent\textbf{Fully Homomorphic Encryption (FHE).}
FHE is an encryption method that permits the execution of computations on encrypted data, termed ciphertexts, without the necessity to decrypt them. This capability ensures that sensitive information remains encrypted throughout the computation process,  enhancing data privacy. 

In FHE, a plaintext space is typically structured as a polynomial ring, for instance, \( R = \mathbb{Z}[X]/(X^N + 1) \), where \( N \) is a power of 2. This polynomial ring can encode more than one value at a time by assigning different coefficients of the polynomial to different data values. The concept of slots arises from this encoding: each coefficient in the polynomial represents a separate ``slot,'' allowing multiple values (e.g., elements of a vector) to be encrypted simultaneously within the same ciphertext.

During the encryption process, a vector \( \vec{v} = (v_1, v_2, \ldots, v_k) \) is encoded into a polynomial where each element \( v_i \) is placed into a distinct slot. For example, if \( \vec{v} \) is encoded into the polynomial \( p(x) = v_1 + v_2x + \ldots + v_kx^{k-1} \), the entire vector can be encrypted at once as a single ciphertext. Operations on this ciphertext, such as addition or multiplication, are performed component-wise in a parallel fashion across all slots. This means that a single homomorphic operation on the ciphertext results in the corresponding operation being applied to each element of the vector simultaneously, known as SIMD (Single Instruction, Multiple Data) operations within FHE.

The main FHE primitives are:
\begin{itemize}
    \item \textbf{FHE Encoding}(\( x \)) \( \to p \): Converts a message \( x \) into a polynomial plaintext format \( p \) that can be processed by the FHE system.
    
    \item \textbf{FHE Key Generation}($1^\lambda$) \( \to (sk, pk) \): Generates a set of keys for the FHE scheme. Here, \( \lambda \) is the security parameter that dictates the strength of the keys. This function outputs a secret key \( sk \), and a public key \( pk \).
    
    \item \textbf{FHE Encryption}(p; \( pk \)) \( \to c \): Encrypts the plaintext \( p \) using the public key \( pk \) to produce the ciphertext \( c \).
    
    \item \textbf{FHE Decryption}(c; \( sk \)) \( \to x \): Decrypts the ciphertext \( c \) using the secret key \( sk \) to recover the original message \( x \).
    
    \item \textbf{FHE Addition}(c, \( \hat{c} \)): Performs an addition operation on ciphertexts \( c \) and \( \hat{c} \), which corresponds to the addition of their respective plaintexts. This operation does not require the evaluation key.
    
    \item \textbf{FHE Multiplication}(c, \( \hat{c} \); \( pk \)): Conducts a multiplication operation on ciphertexts \( c \) and \( \hat{c} \), corresponding to the multiplication of their respective plaintexts. This operation typically requires the evaluation key \( pk \) to process the multiplication homomorphically.
\end{itemize}

Non-linear operations are not supported natively in FHE. Hence, a non-linear activation function in a DNN may be approximated by a polynomial function \cite{cryptonet1,cryptonet2, cryptonet3} or a lookup table \cite{lut}.  

\subsection{Related Works} 

Building upon the foundational concepts introduced in the previous subsection, it is essential to also review the performance and overhead considerations before proposing our method in Section~\ref{sec:design}, as high overheads are often impediment to deployment in real systems. Our evaluation of different FHE verification methods is guided by several critical factors that directly influence their practical applicability and effectiveness:


\noindent\textbf{Overheads.} 
Overheads include space overheads (addition of metadata or magnification of data relative to the original data) and computation overheads (addition of execution time  introduced by the verifiablity). These costs may be incurred both at the server and the client side. It is crucial for a verifiable scheme to keep space and computation overheads low on both ends, especially at the client, where the initial encryption and encoding take place.

\noindent\textbf{Scalability.} 
Ideally, the scheme should scale effectively with the computation size, meaning the overhead relative to the computation size should remain stable or, ideally, diminish. This involves efficient handling of encryption and decryption at the client and managing the computational burden in the server. 

\noindent\textbf{Security.} 
Verifiability should extend beyond just trusted servers; it must also guard against malicious servers with the capability and intent to tamper with data and computations.

\newcommand*\feature[1]{\ifcase#1 -\or\LEFTcircle\or\CIRCLE\fi}
\newcommand*\f[1]{\feature#1}
\makeatletter
\newcommand*\ex[4]{#1&\f#2&\f#3&\f#4}
\begin{table}[ht]
\centering
\footnotesize
\begin{threeparttable}
\caption{Comparison of current FHE verifiablity schemes and our DataSeal method.}
\label{tab:features}
\setlength{\tabcolsep}{5pt}
\begin{tabular}{lccc}
\toprule
\textbf{Scheme} & \textbf{Scalability} & \textbf{Security} & \textbf{Low Overhead} \\
\midrule
\ex{FHE+ABFT~\cite{Awadallah2021, cryptoeprint:2023/231}}{2}{0}{2}\\
\ex{FHE+MAC~\cite{chatel2022verifiable}}{0}{2}{0}\\
\ex{FHE+ZKP ~\cite{viand2023verifiable}}{0}{2}{0}\\
\ex{FHE+TEE~\cite{natarajan2021chex, viand2023verifiable} }{2}{2}{0}\\
\ex{\textbf{DataSeal}}{2}{2}{2}\\
\bottomrule
\end{tabular}
\begin{tablenotes}
\begin{small}
\vspace{0.1in}
\item $\feature2=\text{provide property}$;
$\text{\feature0}=\text{not provide property}$;
\end{small}
\end{tablenotes}
\end{threeparttable}
\end{table}

In Table \ref{tab:features}, we compare our DataSeal method with existing FHE integrity schemes, highlighting key differences in scalability, security, and overhead. FHE+MAC combines message authentication codes with FHE to ensure data integrity but often suffers from scalability issues and high overhead \cite{chatel2022verifiable}. FHE+ZKP uses zero-knowledge proofs to verify computation correctness without revealing underlying data, providing strong security but also at a significant computational cost \cite{viand2023verifiable}. FHE+TEE leverages Trusted Execution Environments to enhance security, offering better scalability but still limited in reducing computational overhead \cite{natarajan2021chex, viand2023verifiable}. In contrast, as we will demonstrate, our DataSeal approach maintains high scalability and security while significantly minimizing overheads. 

\begin{figure}[ht]
    \centering
    \hspace{-0.5cm}
    \begin{tikzpicture}[scale=0.95]
\begin{axis}[
    xlabel={$n$ (squared-matrix size)},
    ylabel={Time in Seconds},
    legend pos=south east,
    grid=both,
    ymode=log,
    log basis y={10},
    ymax=10000
]
\addplot table {
n   FHE-only
1 0.007317
2 0.013145
3 0.028241
4 0.038275
5 0.059206
6 0.070476
7 0.102446
8 0.117411
15 0.372756
16 0.403254
23 0.815831
24 0.889485
31 1.511596
32 1.483812
63 5.648339
64 5.786344
};
\addlegendentry{FHE-only}

\addplot table {
n   FHE+TEE
1	0.001850825
2	0.007028951
3	0.02573094
4	0.035254825
5	0.068667673
6	0.086054113
7	0.143496646
8	0.167906306
15	0.632765274
16	0.651516493
23	1.491599519
24	1.600757154
31	2.641528615
32	2.722580376
63	11.02929701
64	11.03712042
};

\addlegendentry{FHE+TEE}

\addplot table {
n   FHE+MAC
1 0.017121
2 0.039324
3 0.079246
4 0.138299
5 0.217848
6 0.320366
7 0.467224
8 0.608021
15 2.62271
16 3.01912
23 7.47015
24 8.3456
31 16.02885
32 17.24404
63 106.3151
64 110.6386
};
\addlegendentry{FHE+MAC}

\addplot[mark=diamond*] table {
n   FHE+ZK
1	0.28
2	0.45
3	2.9
4	4.312174
5	6.077391
6	8.296522
7	10.81826
8	15.20609
15	63.04348
16	75.65217
23	199.2174
24	211.8261
31	400.9565
32	476.6087
63	2958
64	3321.13
};
\addlegendentry{FHE+ZKP}



\end{axis}
\end{tikzpicture}
    \caption{Performance degradation of current state-of-the-art FHE verifiability schemes. (Setup time and Compute time). Refer to Section~\ref{sec:setup} for the platform that was used.}
    \label{fig:mot}
\end{figure}
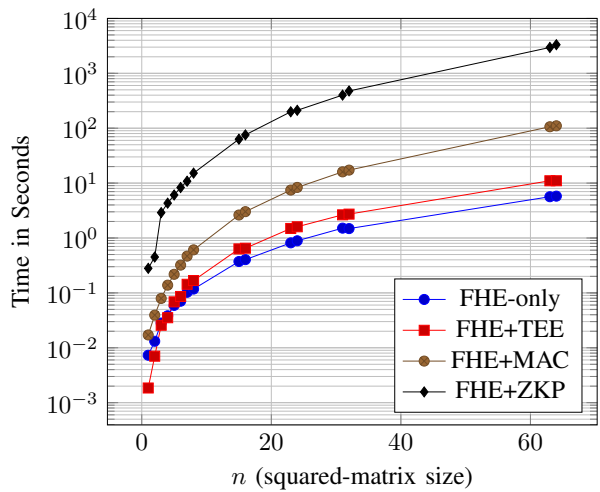

The analysis in Figure~\ref{fig:mot} provides insights into the performance of various Full Homomorphic Encryption (FHE) schemes, focusing on time and memory requirements during client Encoding, server Operation, and client Decoding phases. Unlike ZKP-based schemes, which introduce additional computational steps, the overhead in Message Authentication Code (MAC) based schemes stems from a different process.


\noindent\textbf{FHE+ZKP.}
ZKP-based schemes~\cite{ganesh2021rinocchio} are built on ZKP backends such as Pinicchio~\cite{parno2016pinocchio} and Groth16~\cite{groth2016sgroth16}. However, as shown in Figure \ref{fig:mot}, they typically pose heavy computation overheads, primarily resulting from additional calculations at both the client and the server, e.g., the key generation, proof generation, and proof verification. Firstly, the key generation in the setup phase of ZKP-based schemes is typically computationally intensive, which alone can take as many as $30\times$ runtime compared with the FHE-only scheme through all phases. Moreover, the server Operation phase is also hindered by the heavy computation in proof generation, which makes the server operation time $\sim600\times$ longer. Most importantly, the proof verification over ciphertext also introduces heavy overhead to the Client Decoding phase, where the client spends $\sim400 \times$ more time decoding and verifying the proof than FHE-only schemes. 

\noindent\textbf{FHE+MAC.}
The MAC-based approach in homomorphic encryption, exemplified by systems like VERITAS \cite{chatel2022verifiable}, employs Message Authentication Codes (MACs) integrated with ring learning with error schemes to safeguard both the integrity and privacy of data in server computations. In this methodology, each scalar value of an input vector is transformed into an extended vector that not only encapsulates the original data value but also incorporates challenge values derived from a pseudorandom function. Additionally, the data is replicated multiple times within each vector to enhance security against tampering, hence incurring high space overheads. Given that FHE already incurs high space overheads, additional data magnifications makes deployment highly problematic. 

These extended vectors are then concatenated, forming a robust structure that enables the detection of any data alteration during or after processing. This replication also leads to increased computational overhead. Figure~\ref{fig:mot} shows that FHE+MAC incurs more than an order of magnitude over FHE-only.  

\noindent\textbf{FHE+TEE.}
Recent studies such as \cite{natarajan2021chex, viand2023verifiable} discuss the use of Trusted Execution Environments (TEEs), which offer scalability and strong confidentiality for non-FHE computations but have limitations in integrity protection and performance when applied to FHE. 

However, TEEs cannot be relied upon for full verification. Many TEEs do not even support full integrity protection. For example, AMD SEV~\cite{AMD_SEV} (through its SnP feature) and Intel Scalable SGX~\cite{Intel_SGX} provide isolation of one enclave from others and from the hypervisor by enforcing the ownership of pages. However, memory tampering can occur without having an ownership of a page, and can even occur without using load/store instruction to the target address. For example, in RowHammer attacks, an enclave data may be altered by another party without that party having direct access to data. An earlier version of SGX did support integrity protection via an integrity tree~\cite{intel2021sgx}, but the support was removed in Scalable SGX to avoid overheads that an integrity tree incurs. 

In addition, verifiability goes well beyond integrity protection and supporting it with TEE is challenging. In order for the client to verify computation performed by the server, the client may need an attestation report to ensure the code executed by the server is the correct one. However, that implies that the client needs to trust the enclave in the server to provide the correct expected measurement. Also, TEE model places the trust boundary on the processor. The combination of these factors mean that FHE on TEE require a larger trust base than what traditional FHE computation requires. 

Furthermore, TEE support is not universal and not hardware-agnostic; it is widely applicable on CPUs, but platforms such as NPUs and many GPUs do not support TEEs. A recent NVIDIA GPU (Hopper architecture) supports an enclave model but broadens the trust boundary to the package beyond just the GPU die. 

Finally, TEEs still incur significant performance drawbacks. They provide confidentiality by encrypting data in memory, which is redundant and unnecessary when data is already encrypted with FHE. 

Additionally, as illustrated in Figure~\ref{fig:mot}, running FHE within a TEE, such as AMD SEV which lacks a comprehensive integrity tree, can slow down the process by nearly 2 times. This showcases the practical limitations of TEEs, especially regarding the drastic reduction in performance and the limited scope of verification they provide.

In contrast, by adapting ABFT and securing it, DataSeal offers a more adaptable and efficient solution than TEE. DataSeal provides full verifiability and is hardware-agnostic, hence it can be deployed in any hardware platform. By building on ABFT, DataSeal also does not incur high-performance penalties. However, unlike ABFT, DataSeal utilizes a client-specific secret key, making it secure against data or computation modifications. DataSeal also avoids redundant data encryption that TEE incurs.

\noindent\textbf{FHE+ABFT.}
The FHE+ABFT (Fully Homomorphic Encryption combined with Algorithm-Based Fault Tolerance) scheme offers an innovative approach to providing integrity protection and verification for encrypted data computations. By embedding error detection and correction directly within the computational algorithms, FHE+ABFT significantly reduces the computational overhead associated with these processes, making the scheme highly scalable and suitable for large-scale applications. However, despite these advantages, FHE+ABFT falls short in security against more sophisticated threats. 

One notable vulnerability is its susceptibility to attacks by adversaries with access to encryption oracles, a common scenario in client-server computations. Such adversaries can exploit this access to manipulate encrypted data and the associated checksums. By replacing the entire dataset and its corresponding checksum with fraudulent versions, an adversary can bypass the integrity checks that FHE+ABFT is supposed to enforce. This ability to tamper with undetected data severely undermines the scheme's security, making it inadequate in environments where robust protection against active threats is required. Therefore, while FHE+ABFT excels in efficiency and scalability, its security vulnerabilities necessitate further enhancements or additional protective measures to safeguard against malicious manipulations.

\section{DataSeal Design}\label{sec:design}
\subsection{Correctness Requirements}

While FHE allows for computation on encrypted data, it does not inherently provide mechanisms for verification of the computations performed. This limitation means that while data remains confidential, the correctness and authenticity of the computation cannot be independently verified under standard FHE protocols. On the other hand, ABFT enhances system reliability by embedding verification capability into computational algorithms, but cannot protect against malicious manipulations by adversaries.

DataSeal aims to combine build on ABFT and integrate it to FHE. In secure computation, three critical properties must be ensured: correctness, completeness, and soundness. We follow the definition of them from~\cite{viand2023verifiable}. 
Correctness ensures that an honest computation yields correct decrypted results. Completeness guarantees that the verification computation always accepts an honest computation result. Soundness ensures an adversary cannot make the verifier accept an incorrect or altered computation. While ABFT and FHE aim to achieve these properties to some extent, their limitations can be exploited under certain conditions. 

\begin{definition}[Correctness]
The scheme is correct if an honest computation yields correct decrypted results.
\begin{equation}
\scriptsize
    \Pr\left[
    \text{Dec}_{sk}(c_y) = f(x)
    \;\middle|\;
    \begin{aligned}
        &(pk, sk) \leftarrow \text{KGen}(1^\lambda,f), \\
        &(c_x, \tau_x) \leftarrow \text{Enc}_{\text{key}}(x), \\
        &(c_y, \tau_y) \leftarrow \text{Eval}_{\text{key}}(c_x)
    \end{aligned}
    \right] = 1
\end{equation}
\end{definition}

\begin{definition}[Completeness]
The scheme is complete if the verification computation always accepts an honest computation result.
\begin{equation}
\scriptsize
    \Pr\left[
    \text{Verify}_{sk}(c_y, \tau_x, \tau_y) = 1
    \;\middle|\;
    \begin{aligned}
        &(pk, sk) \leftarrow \text{KGen}(1^\lambda,f), \\
        &(c_x, \tau_x) \leftarrow \text{Enc}_{\text{key}}(x), \\
        &(c_y, \tau_y) \leftarrow \text{Eval}_{\text{key}}(c_x)
    \end{aligned}
    \right] = 1
\end{equation}
\end{definition}

\begin{definition}[Soundness]
The scheme is sound if an adversary cannot make the verifier accept an incorrect or altered computation.
\begin{equation}
\scriptsize
\Pr\left[
\begin{array}{c}
    \text{Verify}_{sk}(c_y, \tau_x, \tau_y) = 1 \\
    \land \\
    \text{Dec}_{sk}(c_y) \neq f(x)
\end{array}
\;\middle|\;
\begin{aligned}
    &(pk, sk) \leftarrow \text{KGen}(1^\lambda,f), \\
    &x \leftarrow A^{\text{OEnc},\text{ODec}}(pk), \\
    &(c_x, \tau_x) \leftarrow \text{Enc}_{\text{key}}(x), \\
    &(c_y, \tau_y) \leftarrow A^{\text{OEnc},\text{ODec}}(c_x, \tau_x)
\end{aligned}
\right] = 1
\end{equation}
\end{definition}

ABFT is not secure because it does not fully meet the completeness and soundness properties. Completeness refers to the ability of the system to correctly verify genuine, unaltered results, which ABFT can achieve. However, soundness, which ensures incorrect computations are detected and rejected, is not fully met. A primary weakness of ABFT is that it relies on a function for computing the checksum that can be inverted, i.e. knowing a value of a checksum that evades verification, the attacker can figure out what alterations to data need to be performed.  An adversary can manipulate the results so that the checksum validations pass, creating a false sense of correctness. This lack of robust soundness against adversarial manipulations is the core weakness that allows such attacks to succeed in ABFT and ABFT combined with FHE.

To address these gaps, this paper proposes an enhanced ABFT approach specifically tailored for integration with FHE systems. Our method replaces the checksum function used in ABFT with a secret checksum function that relies on a client-specific secret key. Unless the attackers know this key, they cannot determine what data should be modified to evade the verification pass. This ensures data privacy through FHE and introduces additional mechanisms to provide robust verification and integrity checks. By strengthening the soundness property, our enhanced ABFT approach can detect and reject malicious manipulations, thereby preventing adversaries from passing incorrect or altered computations as correct. This integration ensures that the system meets the critical properties of correctness, completeness, and soundness, thereby enhancing computations' overall security and reliability in adversarial environments. The specific details on how our proposed method addresses these vulnerabilities and the mitigation strategies are discussed in Section~\ref{sec:design}.

\begin{figure}[t]
    \centering
    \includegraphics[width=0.47\textwidth]{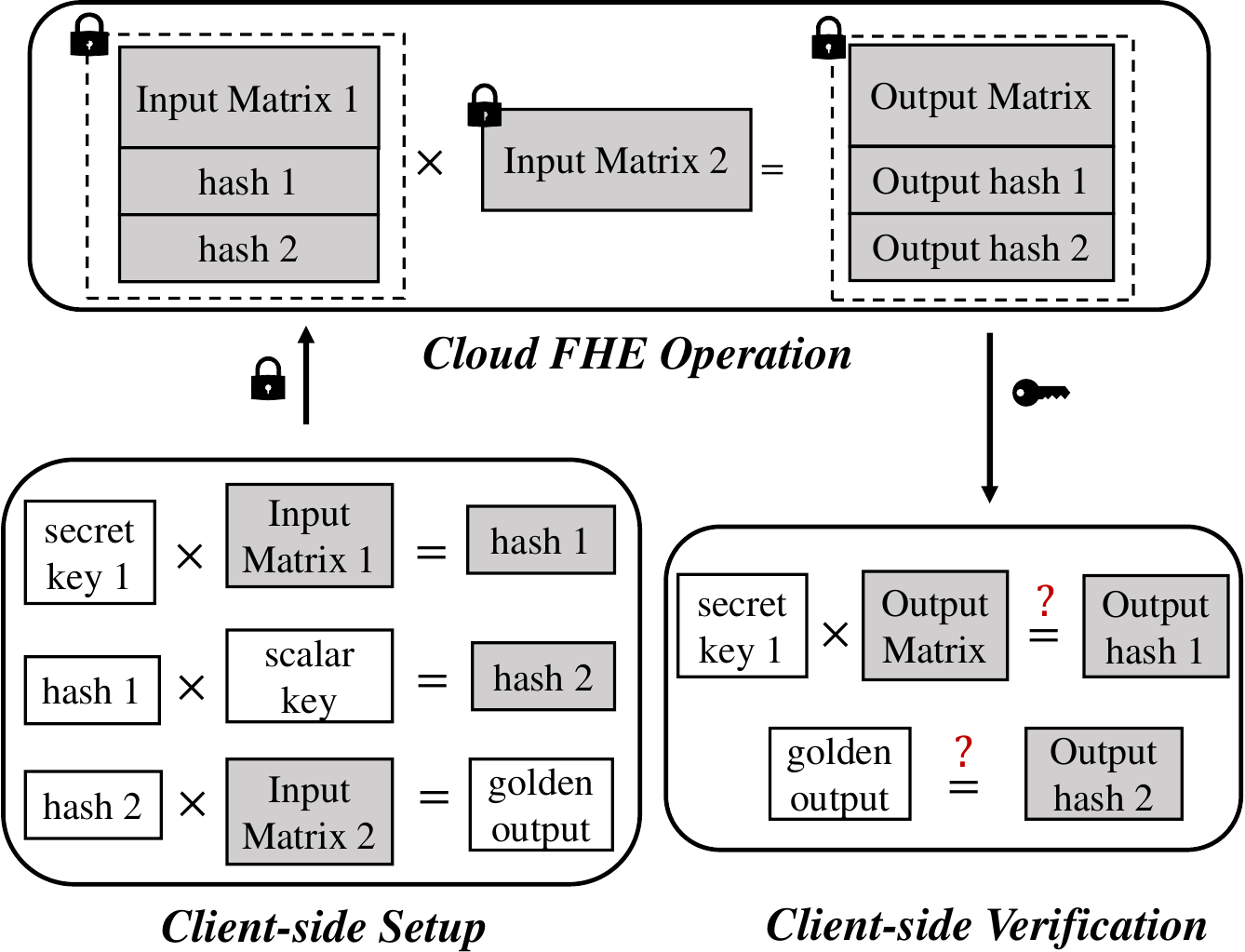}
    \caption{An example of matrix multiplication workflow using our DataSeal. }
    \label{fig:overview}
\end{figure}

\subsection{DataSeal Overview}

Similar to ABFT, DataSeal incurs low space and computation overheads. Whereas ABFT relies on adding one row, DataSeal adds up to two additional rows (depending on computation operation) to the original matrix, regardless of the matrix size. Hence, the space overheads scales with $\frac{2N}{N^2}=\frac{2}{N}$ which asymptotically approaches zero as $N$ increases. Being software-based, DataSeal is hardware-agnostic, making it deployable in various hardware platforms. 

Our DataSeal's workflow is illustrated in Figure~\ref{fig:overview}, showcasing the matrix multiplication process between input matrix 1 and input matrix 2 as an example. Prior to performing FHE computation in the cloud, the client performs an initial encoding setup. In this step, a secret key 1 vector is chosen and multiplied with input matrix 1 to generate a hash value, essentially a weighted checksum. This hash 1 is then scaled to create hash 2, combined with input matrix 2 to produce the golden output. Before server sharing, input matrices 1 and 2 and hash 1 and 2 are encrypted. The server then calculates the FHE matrix to obtain the output matrix and the corresponding output hashes 1 and 2. Post-calculation, these values are decrypted for verification purposes. During the verification phase, the client employs the secret key to multiply with the output matrix, verifying if the result matches output hash 1. Additionally, the golden output is checked against output hash 2. The computation is verified only if both these comparisons are successful, confirming the integrity and accuracy of the computation. While matrix multiplication is used as an illustrative example here, DataSeal's methodology is equally applicable to other matrix operations such as addition, squaring, and more, which will be covered in subsequent sections.

Note that computationally, the client-side setup and verification are much cheaper than matrix multiplication. Hash 1, 2, and golden output are single-dimension vectors. Their multiplication with input matrices 1 and 2 is thus less complex than the matrix multiplication. That is, if the size of matrices 1 and 2 are $N\times N$, the asymptotic computation complexity of the setup and verification, compared to the matrix multiplication, is $\mathcal{O}(\frac{1}{N})$, which again decreases the overheads as the matrix size increases.  

\noindent\textbf{Weighted Checksum.} The idea behind the weighted checksum is to improve security by using a weighted sum instead of a simple checksum. A weighted checksum is generated by multiplying the secret key vector with the input matrix, creating a more secure hash value. Specifically, consider the input matrix \( \mathbf{A} \in \mathbb{R}^{m \times n} \) and the secret key vector \( \mathbf{k} \in \mathbb{R}^n \):

\[
\footnotesize
\mathbf{A} = \begin{pmatrix}
a_{11} & a_{12} & \cdots & a_{1n} \\
a_{21} & a_{22} & \cdots & a_{2n} \\
\vdots & \vdots & \ddots & \vdots \\
a_{m1} & a_{m2} & \cdots & a_{mn}
\end{pmatrix}, \quad
\mathbf{k} = \begin{pmatrix}
k_1 & k_2 & \cdots & k_n
\end{pmatrix}
\]

The weighted checksum \( \mathbf{w} \) can be computed as:
\[
\footnotesize
\mathbf{w} = \mathbf{k} \cdot \mathbf{A} 
=
\begin{pmatrix}
\sum_{i=1}^n k_i a_{i1} & \sum_{i=1}^n k_i a_{i2} & \cdots & \sum_{i=1}^n k_i a_{im}
\end{pmatrix}
\]

Each element in the weighted checksum vector \( \mathbf{w} \) depends on all the secret values of secret vector $\mathbf{k}$. Not knowing $\mathbf{k}$, if the attacker modifies an element of the original matrix $\mathbf{A}$, he/she cannot figure out what checksum value can evade verification check unless he/she knows $\mathbf{k}$ which is a client's secret. Furthermore, the attacker cannot use a linear solver to solve for $\mathbf{k}$ given a system of linear equations with $n$ equations and $m$ unknown variables because the weighted checksum $\mathbf{w}$ is also encrypted with FHE prior to being sent to the server. 

There is only one vulnerability that the attacker can exploit. He/she can multiply the entire matrix by a constant and still pass the verification, as the proportionality of the weights would remain unchanged. Specifically, the adversary can multiply the entire matrix \( \mathbf{A} \) by a constant \( c \), producing a new matrix \( \mathbf{A}' = c \cdot \mathbf{A} \). The weighted checksum for \( \mathbf{A}' \) becomes:

\[
\mathbf{w}' = \mathbf{k} \cdot \mathbf{A}' = \mathbf{k} \cdot (c \cdot \mathbf{A}) = c \cdot (\mathbf{k} \cdot \mathbf{A}) = c \cdot \mathbf{w}
\]

Since the proportionality remains unchanged, the altered matrix \( \mathbf{A}' \) could still pass the verification based on the weighted checksum, revealing a limitation in this approach. In practice, entire-matrix constant scaling is quite restrictive as an attack and may not lead to any useful exploit. However, we would still like to be able to detect such tampering, which leads us to the next scheme. 

\noindent\textbf{Golden Input and Output.} The golden input concept addresses the weighted checksum's limitations. Instead of leaving the weighted checksum as is, it is multiplied by a random constant, treating it as a vector input to the operation. Let \( r \) be the random constant. The golden input \( \mathbf{g} \) is computed as:

\[
\mathbf{g} = r \cdot \mathbf{w} = r \cdot (\mathbf{k} \cdot \mathbf{A})
\]
This golden input \( \mathbf{g} \) is then combined with another input matrix \( \mathbf{B} \in \mathbb{R}^{n \times p} \) to produce a golden output \( \mathbf{R} \):

\[
\mathbf{R} = \mathbf{g} \cdot \mathbf{B} = (r \cdot (\mathbf{k} \cdot \mathbf{A})) \cdot \mathbf{B}
\]

During verification, if an adversary attempts to multiply the entire matrix \( \mathbf{A} \) by a constant \( c \), the new matrix \( \mathbf{A}' \) is:

\[
\mathbf{A}' = c \cdot \mathbf{A}
\]

The weighted checksum for \( \mathbf{A}' \) would be:

\[
\mathbf{w}' = \mathbf{k} \cdot \mathbf{A}' = \mathbf{k} \cdot (c \cdot \mathbf{A}) = c \cdot (\mathbf{k} \cdot \mathbf{A}) = c \cdot \mathbf{w}
\]

The golden input for \( \mathbf{A}' \) would be:

\[
\mathbf{g}' = r \cdot \mathbf{w}' = r \cdot (c \cdot \mathbf{w}) = c \cdot (r \cdot \mathbf{w}) = c \cdot \mathbf{g}
\]

The golden output with the altered matrix \( \mathbf{A}' \) would be:

\[
\mathbf{R}' = \mathbf{g}' \cdot \mathbf{B} = (c \cdot \mathbf{g}) \cdot \mathbf{B} = c \cdot (\mathbf{g} \cdot \mathbf{B}) = c \cdot \mathbf{R}
\]

This would result in a mismatch during verification because \( c \cdot \mathbf{R} \neq \mathbf{R} \), thus failing the verification.

However, relying solely on the golden output is also insufficient. An adversary could calculate the golden input honestly and perform the computation, then overwrite the main data post-computation to pass the verification. Therefore, both the weighted checksum and the golden output are used together to enhance security, ensuring that both must be validated to confirm the integrity and accuracy of the computation.

The overhead introduced by DataSeal is low due to its efficient design. During the encoding phase, the overhead consists of two vector and matrix multiplications to generate the weighted checksum and golden output rows. In the cloud computation phase, the overhead is the addition of two extra rows to the matrix. During the verification phase, the overhead is one vector-matrix multiplication to validate the results. By keeping the computational complexity low through vector-based operations for encoding and verification, DataSeal ensures that the performance impact on the system is negligible, thereby maintaining overall computational efficiency while enhancing security.

\subsection{Message Encoding in DataSeal}
\label{s:setup}

Given two input matrices, \( M_A \) and \( M_B \), the function \( f(M_A, M_B) = M_o \) represents the computation that can be matrix multiplication, addition, or polynomial operations such as squaring. The primary objective of the encoder within \textit{DataSeal} is to transform \( M_A \) and \( M_B \) in such a way that the output \( M_o \) not only adheres to the computational requirements but is also verifiable post-computation. To achieve this, the \textit{DataSeal} encoder, invoked as \texttt{DataSeal.encode(f(.), M\_a, M\_b)}, outputs encoded versions of the matrices \( M^*_A \) and \( M^*_B \), along with a verification key (vk) and a vector \( v_o \) that serves as the golden output for verification purposes. The verification key \( vk \) is generated using a Pseudo-Random Function (PRF), ensuring that each encoding session is unique and secure. This ABFT-based approach enhances the encoder’s ability to maintain integrity and ensure verifiability of the computational results. The subsequent parts will delve into the specifics of the encoding techniques tailored for each type of operation---matrix multiplication, addition, and squaring.

\begin{algorithm}
\caption{\texttt{DataSeal.encode(f(X), M\_A, M\_B)}}
\label{alg:enc_multiplication}
\begin{algorithmic}[1]
\STATE $vk, \alpha = PRF()$
\STATE Compute $v_A = vk \cdotp M_A$.
\STATE Compute $v_B = \alpha \cdotp v_A$.
\STATE Compute $v_o = v_B \cdotp M_B$.
\STATE Append $v_A$ and $v_B$ as the last row of matrix $M_A$, forming $M^*_A$.
\end{algorithmic}
\end{algorithm}

\noindent\textbf{Multiplication.}
For matrix multiplication, the encoding process is defined in Algorithm \ref{alg:enc_multiplication}, which details the steps to prepare Matrix \( A \) owned by the client for multiplication with Matrix \( B \), assumed to be known by both the client and the server. The process begins by generating a verification key (\( vk \)) and a scalar (\( \alpha \)) using a pseudorandom function (PRF). The client computes a blinded version of Matrix \( A \), \( v_A \), by multiplying \( vk \) with \( M_A \), and then applies \( \alpha \) to \( v_A \) to obtain \( v_B \). Subsequently, \( v_B \) is multiplied by Matrix \( B \) to produce the golden output \( v_o \). Finally, \( v_A \) and the result of \( v_B \cdot M_B \) are appended as the last row to Matrix \( A \), forming \( M^*_A \), which is then encrypted using FHE to produce \( \text{enc}_{\text{FHE}}(M^*_A) \). This encoded matrix \( M^*_A \) maintains confidentiality and allows for the verification of the computation's correctness and integrity.

\begin{algorithm}
\caption{\texttt{DataSeal.encode(f(+), M\_A, M\_B)}}
\label{alg:enc_addition}
\begin{algorithmic}[1]
\STATE $vk = PRF()$
\STATE Compute $v_A = vk \cdotp M_A$.
\STATE Compute $v_B = vk \cdotp M_B$.
\STATE Compute $v_o = v_A + v_B$.
\STATE Append $v_A$ as the last row of matrix $M_A$, forming $M^*_A$.
\STATE Append $v_B$ as the last row of matrix $M_B$, forming $M^*_B$.
\end{algorithmic}
\end{algorithm}

\noindent\textbf{Addition.}
The encoding process is outlined in Algorithm \ref{alg:enc_addition} for matrix addition. This process starts with the generation of a verification key (\( vk \)) using a pseudorandom function (PRF). The client then computes \( v_a \) by multiplying \( vk \) with Matrix \( M_A \) and \( v_B \) by multiplying \( vk \) with Matrix \( M_B \). These operations produce vectors \( v_A \) and \( v_B \), which are then added together to form \( v_o \), representing the encoded result of the matrix addition. To ensure the integrity and verifiability of the computation, \( v_A \) is appended as the last row to Matrix \( M_A \), forming \( M^*_A \), and \( v_B \) is appended as the last row to Matrix \( M_B \), forming \( M^*_B \). The augmented matrices \( M^*_A \) and \( M^*_B \) are then encrypted using FHE, enabling secure and verifiable addition operations.

\begin{algorithm}
\caption{\texttt{DataSeal.encode($f((.)^n$),$M_A$)}}
\label{alg:enc_polynomial}
\begin{algorithmic}[1]
\STATE $vk = PRF()$
\STATE For each column $i$ of $M_A$, compute $v_{A_i} = vk_i \times \left( \prod_{j} (M_{A_{ji}}) \right) \mod m$, where $j$ iterates over the rows of $M_A$.
\STATE $v_o = v_{A}^n \mod m $ \COMMENT{Element-wise polynomial}
\STATE Append $v_A$ as the last row of matrix $M_A$ forming $M^*_A$.
\end{algorithmic}
\end{algorithm}

\noindent\textbf{Polynomial Computation.}
Additionally, the encoding for matrix polynomial computation in the DataSeal framework is detailed in Algorithm \ref{alg:enc_polynomial}. This process begins with the generation of a verification key (\( vk \)) using a pseudorandom function (PRF). For each column \( i \) of Matrix \( M_A \), the algorithm computes \( v_{A_i} \) by multiplying \( vk_i \) (a component of \( vk \)) with the product of elements in column \( i \), taken modulo \( m \). This operation is performed element-wise across all columns, where \( j \) iterates over the rows of \( M_A \). After computing these values, each \( v_{A_i} \) is then raised to the power \( n \) and taken modulo \( m \) to achieve the element-wise polynomial values \( v_A \). These computed values \( v_A \) are then appended as the last row of Matrix \( M_A \), forming \( M^*_A \). The augmented matrix \( M^*_A \) is then encrypted and processed in the cloud. Upon decryption, the client performs verification checks to confirm the integrity of the polynomial computation and the matrix data, ensuring that the computations are accurate and secure.

\subsection{FHE Evaluation and DataSeal Verification}
\label{s:cloud}

We combine DataSeal encoding with a Homomorphic Encryption (HE) scheme to develop a homomorphic authenticator. DataSeal encoding occurs on plaintext, so no modification is needed for the FHE algorithm. DataSeal's verification protocol ensures the consistency of the verification key, matrix output, and embedded redundancy, verifying that output vectors \(v_o\) match their precomputed values from the encoding phase. The verification mechanisms are tailored for each operation type (addition, multiplication, and polynomial computations).

\label{s:verification}
\begin{algorithm}
\caption{Matrix Multiplication Evaluation in DataSeal}
\label{alg:ver_multiplication}
\begin{algorithmic}[1]
\STATE Separate $v_1$ (last row of $M_C^*$), $v_2$ (second last row of $M_C^*$), and the rest of $M_C$.
\STATE Check if $v_2$ equals $v_o$.
\STATE Check if $v_k \times M_C$ equals $v_1$.
\end{algorithmic}
\end{algorithm}

\noindent\textbf{Matrix Multiplication Verification:}
In Algorithm \ref{alg:ver_multiplication}, the client decrypts $enc_{FHE}(C^*)$ to obtain $C^*$. The client first verifies the correctness of the cloud's computation by checking if $v_2 = v_o$ and then confirms the integrity of matrix $A$ by verifying if $v_k \times M_C = v_1$. Additionally, the client verifies that the computation result matches the precomputed golden output.

\textit{Correctness:} To verify matrix multiplication \(A \cdot B = C\), apply a weighted checksum vector \(v\) with the verification key \(v_k\) to \(A\), yielding \(v_k A = w\). Multiply \(w\) by \(B\) to get \(w \cdot B = v_k \cdot A \cdot B = v_k \cdot C\). Verifying \(v_k \cdot C = w \cdot B\) confirms the correctness of the multiplication. Additionally, the golden output \(v_o\) provides an extra layer of confirmation, ensuring the final result matches the precomputed expected value.

\textit{Soundness:} If \(v_k \cdot C = w \cdot B\), it ensures that the original multiplication \(A \cdot B = C\) was performed correctly, as the weighted checksum with verification key captures errors in the process. The golden output \(v_o\) further verifies that no tampering occurred during computation.

\textit{Completeness:} If the original multiplication is correct, \(v_k \cdot C = w \cdot B\) will hold, guaranteeing the verification will not produce false negatives. The golden output \(v_o\) ensures that all valid computations are confirmed, reinforcing the integrity of the verification process.

\begin{algorithm}
\caption{Secure Matrix Addition in DataSeal}
\label{alg:ver_addition}
\begin{algorithmic}[1]
\STATE Separate $v_1$ (last row of $M_C^*$) and the rest of $M_C$.
\STATE Check if $v_o$ equals $v_1$.
\STATE Check if $v_k \times M_C$ equals $v_C$.
\end{algorithmic}
\end{algorithm}

\noindent\textbf{Matrix Addition Verification:}
In Algorithm \ref{alg:ver_addition}, the client decrypts \(\text{enc}_A(C^*)\) to get \(C^*\). The client first checks if \(v_o = v_1\) and then verifies the integrity of matrix \(A\) by checking if \(v_k \times M_C = v_C\). Additionally, the client verifies that the computation result matches the precomputed golden output.

\textit{Correctness:} For \(A + B = C\), apply the weighted checksum vector \(v\) with verification key \(v_k\) to \(A\), \(B\), and \(C\). If \(v_k \cdot C = v_k \cdot A + v_k \cdot B\), the addition is correct. The golden output \(v_o\) further confirms that the final result matches the expected value.

\textit{Soundness:} If \(v_k \cdot C = v_k \cdot A + v_k \cdot B\), it ensures that \(A + B = C\) was performed correctly. The weighted checksum with verification key and the golden output \(v_o\) verify the computation's integrity.

\textit{Completeness:} If the original addition is correct, \(v_k \cdot C = v_k \cdot A + v_k \cdot B\) will hold, ensuring no false negatives. The golden output \(v_o\) guarantees that all valid additions are verified accurately.

\begin{algorithm}
\caption{Secure Matrix Polynomial Verification}
\label{alg:ver_polynomial}
\begin{algorithmic}[1]
\STATE Separate $v_C$ (last row of $M_C^*$) and the rest of $M_C$.
\STATE Check if $v_C$ equals $v_o$.
\STATE Check if $v_Ci = v_k i^n \times \left( \prod_{j} ( A_{ji}) \right) \mod m$. \COMMENT{Integrity check of $M_A$ in computation}
\end{algorithmic}
\end{algorithm}

\noindent\textbf{Verification for Polynomial Computation:}
In Algorithm \ref{alg:ver_polynomial}, after decrypting \(\text{enc}_A(C^*)\), the client first verifies if \(v_C\) equals \(v_o\) and then checks if for each column \(i\), \(v_Ci = v_k i^n \times \left( \prod_{j} ( A_{ji}) \right) \mod m\). Additionally, the client verifies that the computation result matches the precomputed golden output.

\textit{Correctness:} For polynomial computations, apply the weighted checksum with verification key to each matrix column. The method ensures multiplicative relationships are maintained in the results, verifying the correctness of complex computations. The golden output \(v_o\) confirms that the final result is as expected.

\textit{Soundness:} If the checksum condition holds, it ensures that the polynomial computation was performed correctly. The verification key and golden output \(v_o\) further validate the computation's integrity.

\textit{Completeness:} If the original computation is correct, the checksum condition will hold, ensuring no false negatives. The golden output \(v_o\) guarantees that all valid polynomial computations are verified accurately.


\subsection{Design Analysis and Limitation}
\label{s:analysis}

\noindent\textbf{Matrix Computation.} 
DataSeal is optimized for matrix operations like multiplication, polynomial, and addition. It introduces scalable overhead with two extra rows, significant for small matrices but less impactful for larger ones, making it efficient for large-scale computations. However, its efficiency drops with non-matrix data types like single values or vectors due to data replication instead of a more efficient compression approach.

\noindent\textbf{One Shot Integrity Protection.} 
DataSeal can verify only one operation at a time, requiring re-encoding for each subsequent operation. This limitation adds overhead in complex tasks. For instance, in \(A \times B + C\), DataSeal validates \(A \times B\) but must re-encode before adding \(C\), increasing overhead in sequential operations.

\noindent\textbf{Large-scale Data Transfer.} 
DataSeal can easily adapt to latency sensitive applications in the cloud computing scenarios.
Verification latency, data volume and bandwidth are not a concern for a single NN layer. For multilayer NN, verification is required per layer, but the latency can be hidden, e.g. verifying layer $x$ is overlapped with inferencing layer $x+1$. Data volume and bandwidth have increased but are still much smaller than alternatives (ZKP and MAC).

However, when considering large-scale data transfers, such as in scenarios with massive datasets, DataSeal's matrix-based encoding introduces overhead that scales with the matrix dimensions, rather than with the overall dataset size. This results in a more predictable and manageable overhead compared to traditional cryptographic techniques (e.g., ZKP or MAC), where the overhead typically scales with the entire data volume. As a result, DataSeal remains efficient even in large-scale cloud computing operations.

\noindent\textbf{Robustness to Attacks.} Under the setting of DataSeal, clients encrypt all inputs, including hashes, restricting adversaries from manipulating ciphertexts, which does not result in valid encrypted hashes. Furthermore, verification keys and scalars are dynamically generated for each session using a PRF, which mitigates replay attacks~\cite{mo2009secure}. DoS attacks~\cite{long2001trends} are a risk but are beyond the goal of DataSeal, which is to enhance integrity checks for outsourced computations.

\section{Experimental Methodology}

\subsection{Experimental Setup}
\label{sec:setup}
Our experimental platform used Google Cloud's n2d-standard-2 (2 cores, 8 GB RAM) with Ubuntu Linux 20.04 LTS. We employed Microsoft SEAL 4.1 \cite{sealcrypto} for homomorphic encryption. The same setup was used for clients, with the addition of AMD SEV TEE for the Confidential VM service.

\subsection{Compared Schemes}
We compared FHE+MAC \cite{chatel2022verifiable}, FHE+ZKP \cite{fhe-zkp}, and FHE+TEE \cite{sev2020strengthening} with our DataSeal across various applications, including matrix operations and neural networks. Additionally, DataSeal was evaluated against AMD SEV TEE \cite{sev2020strengthening} to gauge performance.

\subsection{Workloads}
We assessed DataSeal across three types of computation: kernels, microbenchmarks, and end-to-end neural networks using the BGV method provided by Microsoft SEAL 4.1 \cite{sealcrypto}.

\noindent\textbf{Matrix Operations.} We evaluated {\em Matrix Multiplications} and {\em Matrix Addition}, commonly used in machine learning and HPC. Matrix addition reveals DataSeal's overhead relative to computation, while matrix multiplication shows how overhead scales with input size. 

\noindent\textbf{Neural Network Micro-benchmarks.} We evaluated linear operation {\em Convolution}, used for feature extraction in neural networks. For {\em Polynomial Activation}, we use square activation for FHE encrypted DNN \cite{ranaldi2022crypto}.

\begin{figure*}[ht]
\centering
    \includegraphics[width=0.9\linewidth]{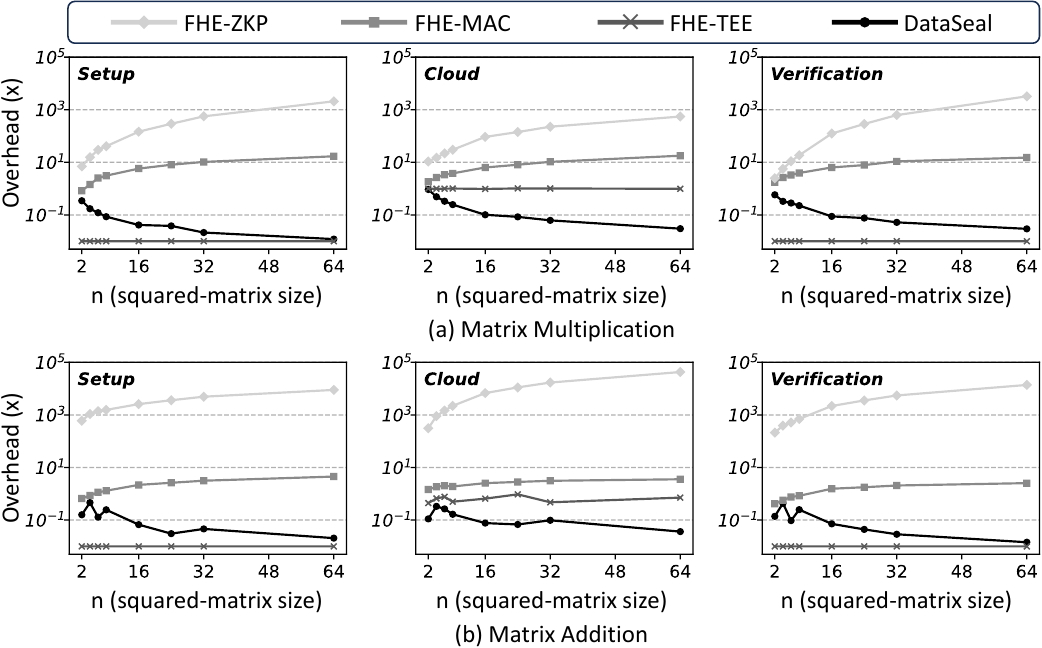}
    \captionsetup{skip=2pt}
    \caption{Overhead comparison of various schemes on matrix operations. The y-axis represents the running overhead, calculated by subtracting the FHE-only running time from each scheme's running time and then dividing this by the FHE-only running time.}
    \label{fig:matrix-overhead-all}
\end{figure*}

\noindent\textbf{End-to-end Neural Networks.} We evaluated LeNet-5 \cite{lecun1998gradient} on the MNIST dataset and AlexNet \cite{krizhevsky2017imagenet} on the CIFAR-10 dataset. LeNet-5 includes convolutional and dense layers, while AlexNet features deep convolutional and fully connected layers. Both are representative models in FHE-based learning \cite{lou2020autoprivacy, CryptoNets:ICML2016, CryptoDL:arxiv17, GAZELLE:USENIX18, al2020towards, xue2024cryptotrain, xue2022audit, lou2019she, Lou2020AutoQ, lou2020falcon, lou2021hemet}.

\section{Result}
\begin{figure*}[ht]
\centering
    \centering
    \includegraphics[width=0.9\linewidth]{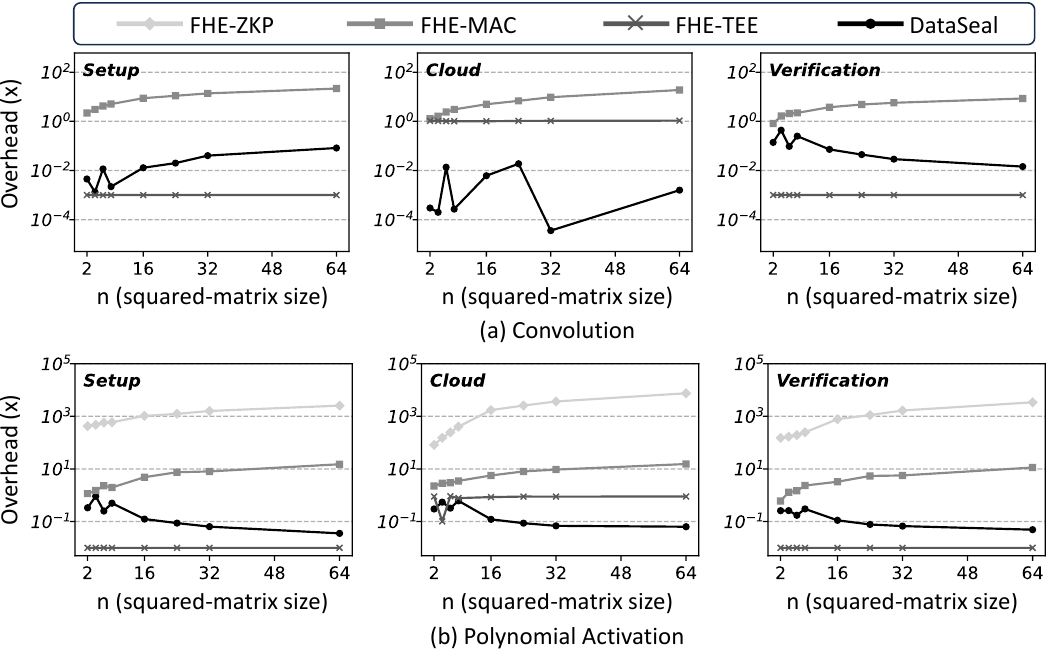}
    \captionsetup{skip=2pt}
    \caption{Overhead comparison of various schemes on neural network micro-benchmark. The y-axis represents the running overhead, calculated by subtracting the FHE-only running time from each scheme's running time and then dividing this by the FHE-only running time. In the case of (a) convolution, FHE-ZKP was not included in the comparison due to its significantly slower performance compared to other methods.}
    \label{fig:dnn-overhead-all}
\end{figure*}

\subsection{Matrix Operations}

\noindent\textbf{Matrix Multiplication.}
Figure \ref{fig:matrix-overhead-all} (a) shows the overheads of DataSeal, FHE-TEE, FHE-ZKP, and FHE-MAC compared to FHE-only performance. DataSeal's constant overhead from two vector-matrix multiplications diminishes relatively as data size increases. In contrast, FHE-MAC's overhead grows with data size due to data replication. FHE-TEE has constant overhead during encoding and verification but incurs significant overhead in computation due to double encryption.

In cloud computation, FHE-TEE's double encryption design doubles the computation time compared to FHE-only. FHE-MAC's and FHE-ZKP's overheads increase with data size due to data replication. DataSeal's overhead decreases with data size, resulting in only a 2.99\% overhead compared to FHE-only at the end of the experiment. This efficiency and scalability make DataSeal suitable for large-scale data tasks.

\noindent\textbf{Matrix Addition.}
Figure \ref{fig:matrix-overhead-all} (b) compares overheads for matrix addition. DataSeal maintains constant overhead with diminishing relative impact as data size grows, while FHE-ZKP and FHE-MAC have increasing overheads due to data replication. FHE-TEE remains constant during encoding and verification but incurs significant overhead in computation.

In cloud computation, FHE-TEE's overhead is consistently high due to double encryption, resulting in a 59.18\% overhead. DataSeal, however, shows a decreasing relative overhead, ending with a 1.85\% overhead compared to FHE-only. This consistent efficiency across both matrix operations positions DataSeal as an effective solution for large-scale data tasks.

\begin{figure*}[ht!]
\centering
    \centering
    \includegraphics[width=0.7\linewidth]{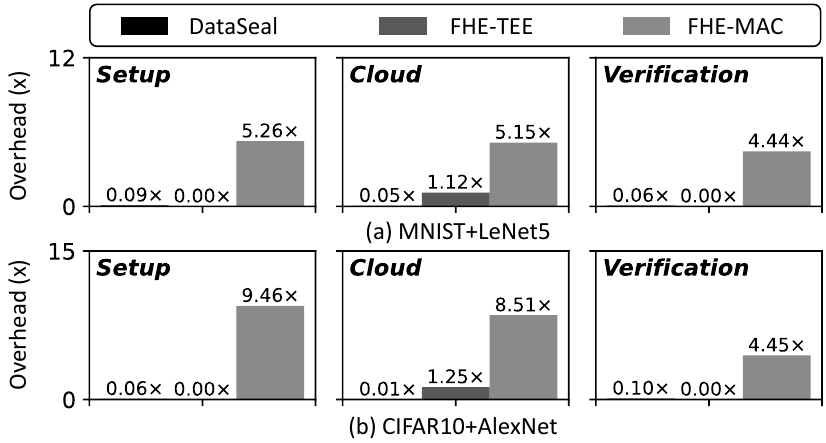}
    \captionsetup{skip=1pt}
    \vspace{0.1in}
    \caption{Overhead comparison among DataSeal, FHE-TEE, and FHE-MAC on neural network inference. The y-axis represents the running overhead, calculated by subtracting the FHE-only running time from each scheme's running time and then dividing this by the FHE-only running time.}
    \label{fig:overhead1}
\end{figure*}

\subsection{Neural Network Micro-benchmark}

\noindent\textbf{Convolution.}
In CNNs, convolution extracts features from input data by sliding a filter across the input and computing the dot product with overlapping regions. This operation, which captures spatial hierarchies in images, can be transformed into a matrix multiplication using the \textit{im2col} method, leveraging optimized routines for faster processing.

Figure \ref{fig:dnn-overhead-all} (a) compares convolution overheads for DataSeal, FHE-TEE, FHE-ZKP, and FHE-MAC. FHE-MAC and FHE-ZKP overheads increase with data size due to data replication. FHE-TEE maintains constant overhead but is significantly higher due to double encryption. DataSeal's setup overhead increases with data size due to converting convolutions into matrix multiplications, yet it achieves almost zero overhead (0.23\%) in cloud environments due to efficient use of FHE slots during conversion.

\noindent\textbf{Polynomial Activation.}
DNNs use activation functions like ReLU, sigmoid, and tanh to introduce non-linearity. In FHE, only addition and multiplication are supported, so polynomial activations, such as squaring, are used.

Figure \ref{fig:dnn-overhead-all} (b) shows the overhead for square activations in FHE. FHE-MAC and FHE-ZKP have increasing overhead with data size due to data replication. FHE-TEE maintains constant but high overhead due to double encryption. DataSeal shows decreasing relative overhead as data size increases, with only 61\% overhead on a $64\times64$ matrix, demonstrating its scalability and efficiency for large data sets in FHE.

\subsection{End-to-End Neural Networks}

Figure \ref{fig:overhead1} compares the computational overhead in Deep Neural Network (DNN) applications among DataSeal, FHE-TEE, and FHE-MAC. The figure shows that DataSeal demonstrates a significantly lower overhead than FHE-TEE and FHE-MAC. FHE-MAC, which uses the slots available in Fully Homomorphic Encryption (FHE) for integrity protection, can mask its latency effectively. This utilization of slots is a strategic approach that takes advantage of the inherent capabilities of FHE, allowing for a certain degree of efficiency in basic operations.

However, as the complexity of the computations increases, the limitations of FHE-MAC become more evident. In complex computational tasks, such as matrix multiplication and convolution operations within DNNs, the FHE-MAC framework is not optimized, leading to a significant increase in computational overhead, e.g., $8.51\times$ slower than FHE-only for a single inference on AlexNet for CIFAR-10. DataSeal, with its unique design and architecture, is more adaptable to these complex computational requirements. It is specifically tailored for matrix operations and offers substantial optimizations in these areas, resulting in a much reduced overhead compared to FHE-MAC. Specifically, DataSeal only has $0.05\times$ and $0.01\times$ computation overheads on LeNet-5 and AlexNet, respectively. 

It is important to note that FHE-TEE maintains a constant overhead throughout different operations, as seen in the previous sections. This consistency in overhead, regardless of the complexity of the operation, distinguishes FHE-TEE from the other frameworks. However, it may not be as efficient as DataSeal in handling complex matrix computations in DNN applications. For instance, FHE-TEE has $1.12\times$ and $1.25\times$ overhead on MNIST-LeNet5 and CIFAR10-AlexNet, respectively.

\section{Conclusion}
To assure the authenticity of private computations on encrypted data, we introduced DataSeal. This system leverages the low overhead of the ABFT with the confidentiality of FHE. This structure reliably validates computations, ensuring efficiency and scalability in client-cloud computing setups. Extensive evaluations in various areas, including matrix operations and neural network analysis, have proven DataSeal's superior performance and scalability on both the client and server fronts. DataSeal consistently maintains minimal overhead, adding only a few extra matrix rows, thereby significantly reducing the verification burden while ensuring data integrity.

\section{Acknowledgement}
This work was supported in part by UCF, ONR grant N00014-23-1-2136, and NSF grant 2106629. 
We thank the anonymous reviewers for their feedback and the shepherd for his/her invaluable guidance throughout the revision process.




\bibliographystyle{plain}
\bibliography{main}

\newpage

\appendices

\section{Meta-Review}
The following meta-review was prepared by the program committee for the 2025 IEEE Symposium on Security and Privacy (S\&P) as part of the review process as detailed in the call for papers.

\subsection{Summary} This paper presents a novel method to verify computations performed using Fully Homomorphic Encryption (FHE), named DataSeal. DataSeal integrates Algorithm-Based Fault Tolerance (ABFT) into FHE to ensure the integrity of computations with low overhead. The method is designed for matrix operations and neural network inference tasks, leveraging secret client-specific checksums for verification. The proposed system is benchmarked against existing solutions like MAC, ZKP, and TEE, showing superior efficiency as the problem size increases.

\subsection{Scientific Contributions} \begin{itemize} \item Creates a New Tool to Enable Future Science \item Provides a Valuable Step Forward in an Established Field \end{itemize}

\subsection{Reasons for Acceptance} \begin{enumerate} \item The verifiability of computations in FHE has been a significant challenge. DataSeal offers an effective solution that addresses this gap with minimal overhead.

\item DataSeal opens new opportunities for researchers working with FHE and matrix operations. Its lightweight design makes it a valuable tool for future applications in secure computing.

\item The work builds on existing methods (ABFT and FHE) but provides significant advancements by integrating these techniques in a new way to enable verifiable encrypted computations. \end{enumerate}

\subsection {Noteworthy Concerns} The authors need to expand their evaluation to other computational workloads and real-world attack scenarios to strengthen the generalizability and robustness of DataSeal.

\section{DataSeal's Formal Security Game}

\noindent1. \textbf{Setup}

\begin{itemize}
    \setlength{\leftmargini}{0pt}
    \item Key Generation. The client generates a secret key $sk$ and public key $pk$ using FHE.
    \item Encoding. The client encodes matrices \( A \) and \( B \), generating encrypted matrices \( M^*_A \) and \( M^*_B \), using a secret checksum \( sk \) and a random scalar \( r \).
    \item Transmission. The client sends the encrypted matrices to the server for computation.
\end{itemize}




\noindent2. \textbf{Adversary Actions}
\begin{itemize}
    \setlength{\leftmargini}{0pt}
    \item Computation. The adversarial server computes \( C \) (e.g., matrix multiplication \( C = A \cdot B \)) and attempts to generate a forged result \( C' \neq A \cdot B \).
    \item Forgery Attempt. The adversary sends the client a forged result \( C' \) and corresponding checksums \( v'_1 \) and \( v'_2 \) (based on the fake result).
\end{itemize}

\noindent3. \textbf{Client Verification}
\begin{itemize}
    \setlength{\leftmargini}{0pt}
    \item The client uses the secret key \( sk \) and scalar \( r \) to recompute the expected checksums \( v_1 \) and \( v_2 \).
    \item The client accepts the result only if: \( v'_1 = \text{weighted checksum of } C' \), \( v'_2 = r \times v'_1 \).
\end{itemize}

\noindent4. \textbf{Winning Condition}

The adversary wins if they can generate a forged result \( C' \) with fake checksums \( v'_1, v'_2 \) that the client accepts.

\noindent5. \textbf{Security Definition}

DataSeal is unforgeable if, for any polynomial-time adversary \( \mathcal{A} \), the probability that the client accepts a forged result \( C' \) with valid checksums is negligible:
    \[
    \Pr[\text{Client accepts } C' \mid C' \neq A \cdot B] \leq \text{negl}(\lambda)
    \]
    where \( \lambda \) is the security parameter (size of \( k \)).

\noindent6. \textbf{Summary}

This game captures the unforgeability of the checksums in DataSeal. Without knowledge of the secret key \( k \) and scalar \( r \), the adversary cannot generate valid checksums for a forged result, ensuring the integrity of the computation.










\end{document}